\documentclass[a4paper]{article}
\usepackage{latexsym}
\usepackage{amsthm,amsmath,amssymb}

\usepackage[dvips]{graphicx, color}
\usepackage[title]{appendix}

\usepackage{color}
\usepackage{bbm}

\newtheorem{thm}{Theorem}

\newtheorem*{theorem-non}{Theorem}

\newtheorem{prop}[thm]{Proposition}

\newtheorem{con}[]{Condition}
\newtheorem{ex}[]{Example}

\newtheorem{cor}[thm]{Corollary}

\newtheorem{lem}[thm]{Lemma}

\newtheorem{defn}[thm]{Definition}

\newcommand{\cE}{{\mathcal E}}
\newcommand{\cF}{{\mathcal F}}

\newcommand{\cM}{{\mathcal M}}

\newcommand{\A}{{\Bbb A}}

\newcommand{\E}{{\Bbb E}}
\newcommand{\F}{{\Bbb F}}

\newcommand{\I}{{\Bbb I}}

\renewcommand{\O}{{\Bbb O}}

\newcommand{\R}{{\Bbb R}}

\newcommand{\Z}{{\Bbb Z}}

\def\O{\mathcal O}

\def\lra{\longrightarrow}

\def\={\:=\:}  \def\+{\,+\,}

\def\a{\alpha}   \def\ba{\overline\a}    
   \def\r{\bold r}

\def\be{\begin{equation}}   \def\ee{\end{equation}}
\def\bes{\begin{equation*}}   \def\ees{\end{equation*}}
\def\ba{\begin{aligned}}   \def\ea{\end{aligned}}
\def\bc{\begin{cases}}   \def\ec{\end{cases}}
\def\bp{\begin{proof}}   \def\ep{\end{proof}}

\newcommand{\res}{\mathrm{res}}

\newcommand{\ord}{\mathrm{ord}}
\newcommand{\Aut}{\mathrm{Aut}}

\def\SL{\mathrm{SL}}

\newcommand{\ov}{\overline}
\def\qqan{\qquad\mathrm{and}\qquad}
\def\qan{\quad\mathrm{and}\quad}

\newcommand{\dis}{\displaystyle}

\def\Ker{\mathrm{Ker}}
\def\GL{\mathrm{GL}}
\def\SL{\mathrm{SL}}

\def\Llra{\Longleftrightarrow}
\def\lra{\longrightarrow}

\def\ov{\overline}

\def\lan{\langle}
\def\ran{\rangle}

\def\l{\ell}
\def\lan{\langle}
\def\ran{\rangle}

\def\bbm1{\mathbbm 1}

\def\whz{\widehat\zeta}

\def\be{\begin{equation}}   \def\ee{\end{equation}}
\def\bes{\begin{equation*}}   \def\ees{\end{equation*}}
\def\bea{\begin{equation}\begin{aligned}}   
\def\eea{\end{aligned}\end{equation}}

\def\wt{\widetilde}

\newcommand{\lhra}{\lhook\joinrel\longrightarrow}

\def\lmt{\longmapsto}

\def\om{\omega}

\def\l{\left}
\def\r{\right}
\def\bm{\begin{matrix}}
\def\em{\end{matrix}}
\def\bpm{\begin{pmatrix}}
\def\epm{\end{pmatrix}}
\def\Hom{\mathrm{Hom}}
\def\wt{\mathrm{wt}}
\def\ev{\mathrm{ev}}
\def\rk{\mathrm{rank}}
\def\diag{\mathrm{diag}}
\def\Om{\Omega}
\def\CFr{C_{F, r}(D, g)}

\begin{document}
\title{\bf  Codes and Stability} 
\author{Lin WENG}  
\date{}
\maketitle

\begin{abstract}
We introduce new yet easily accessible codes for elements of $\GL_r(\A)$ with $\A$ the 
adelic ring of a (dimension one) function field over a finite field. They are linear codes, 
and coincide with classical algebraic geometry codes when $r=1$.  Basic properties of 
these codes are presented. In particular, when offering better bounds for the 
associated dimensions, naturally introduced is the well-known stability condition.  
This condition is further used to determine the minimal distances of these codes. 
To end this paper, for reader's convenience, we add two appendices on some details of 
the adelic theory of curves and classical AG codes, respectively.
\end{abstract}

\section{Introduction}

Let $F/\F_q$ be a function field of an integral regular projective curve $X$ over $\F_q$.
Denote by $\A$ its adelic ring with $\O$ the ring of integers, and, for a positive integer 
$r$, denote by $\GL_r(\A)$ the general linear group with coefficients in $\A$. Fix a 
degree $n$ divisor $D=\sum_{i=1}^np_i$. 

For an element $g=(g_p)_p\in \GL_r(\A)$, we introduce a subspace $\A^r(g)$ of $\A^r$
by
\be
\A^r(g):=\big\{a\in \A^r\,|\, ga\in \O^r\big\},
\ee
and define the comology groups of $g$ on $F$ by
\be
H^0(F,g):=\A^r(g)\cap F^r\qqan H^1(F,g):=\A^r\big/(\A^r(g)+F^r).
\ee
In addition, we introduce the space of $r$-multiple differentials $\Om_F^r(g)$ by
\be
\Om_F^r(g):=\big\{(\om_j)\in \Om_F^r:\,|\, (\phi(\om_j))\in g\O^r\big\}
\ee
where $\Om_F$ denote the space of rational differentials on $X$ and 
$\phi: \Om_F\to F, \om\mapsto \frac{\om}{\om_0}$ is an $F$-isomorphism with $\om_0$ 
a $D$-special rational differential. It is known, see e.g. \S2  that $H^i(F,g)$ satisfy the 
standard topology duality and the Riemann-Roch relation. For example, we have
\be
\Om_F^r(g)\simeq H^1(F,\iota_{(\om_0)}g^{-1})
\ee
where, for a divisor $E$ on $X$, we set 
$\iota_E=\big(\pi_p^{\ord_p(E)}\big)\in \I:=\GL_1(\A)$ to be its characterizing idele.

If $g$ is $D$-balanced, we introduce the following codes
\bea
C_{F, r}(D, g)
:=&\Big\{\big(f_j(p_1)\big),\ldots, \big(f_j(p_n)\big)\,|\,f=(f_j)\in H^0(F,g)\Big\},\\
C_{\Om,r}(D, g):=&\Big\{\big(\om_{j,p_1}(1)\big),\ldots, \big(\om_{j,p_n}(1))\big)\,|\,
(\om_j)\in \Om_F^r(g(-D))\Big\}.
\eea
Here $g(-D):=\iota_D^{-1} g$.
In this paper, we show the following results on basic properties of the above codes.
\begin{theorem-non} Let $g\in \GL_r(\A)$ be  $D$-balanced.
\begin{enumerate}
\item [(0)] If $E$ is a divisor with support away from the $p_i$'s, then
\bes
C_{F, 1}(D, \iota_E)=C_{L}(D, E)\qqan C_{\Om, 1}(D, \iota_E)=C_{\Om}(D, E).
\ees
\item [(1)] $C_{F, r}(D, g)$ and $C_{\Om,r}(D, g)$ are linear codes of length $rn$ and 
mutually dual to each other. That is,
\bes
C_{\Om,r}(D, g)^\perp=C_{F,r}(D, g)\qan 
C_{\Om,r}(D, g)=C_{F,r}(D, \iota_{(\om_0)+D}g^{-1}).
\ees
\item [(2)] The dimensions $k_{D, g}$ and $k_{D, g}^\perp$ of $C_{F, r}(D, g)$ and 
$C_{\Om,r}(D, g)$ are given by
\bes
h^0(F,g)-h^0(F,g(-D))\qqan h^1(F,g(-D))-h^1(F,g),
\ees
respectively. In particular,
\bes
\dim_{\F_q}C_{F, r}(D, g)+\dim_{\F_q}C_{\Om, r}(D, g)\=n\,r.
\ees
\item [(3)]  Assume, in addition, $g$ is semi-stable.
\begin{enumerate}
\item [(a)] For the dimensions of the codes $C_{F, r}(D, g)$ and $C_{\Om,r}(D, g)$,
\begin{enumerate}
\item [(i)] If $\deg (g)<rn$, then
\bes
k_{D,g}=h^0(F, g)\geq \deg(g)-r(g-1).
\ees
\item [(ii)] If $\deg (g)<2r(g-1)$, then
\bes
k_{D,g}^\perp=h^1(F, g(-D))\geq r\big(\deg(D)+(g-1)\big)-\deg(g).
\ees
\item [(iii)] If $rn>\deg(g)>2r(g-1)$, then
\bes
k_{D,g}= \deg(g)-r(g-1)\ \,\mathrm{and}\  \,
k_{D,g}^\perp= r\big(\deg(D)+(g-1)\big)-\deg(g).
\ees
\end{enumerate}
\item [(b)] For the minimal distance $d_{D,g}$ of the code $C_{F, r}(D, g)$, we have
\bes
d_{D,g}\geq nr-\deg(g).
\ees
In particular, if $\deg(g)<rn$, we have
\bes
r(n-(g-1))\leq k_{D, g}+d_{D, g}\leq rn+1.
\ees
\end{enumerate}
\end{enumerate}
\end{theorem-non}
Certainly,  the above results coincides with these for classical algebraic geometry codes,
since, when $r=1$, the stability condition is automatic.

We end this introduction by the following remarks. After completing this paper, as a 
preparation for open lists, we make a searcher on the internet and find the 
paper of V. Savin on \lq Algebraic-geometric codes from vector bundles
and their decoding' at arXiv:0803.1096.  While there are some idea overlaps, our 
codes and approaches are quite different. Also I would  like to thank J. Yamada for his 
explanations of AG codes in our weekly seminar a few days ago, which for the 
first time making AG codes known to me. Immediately, I realized that these codes admit 
natural generalizations using adelic cohomologies we developed several years ago. 
Current work is the outcome of this line of thoughts.

\section{Rank $r$ Codes}
\subsection{Adelic Cohomology Theory}

Let $F/\F_q$ be the function field of an integral regular projective curve $X$ over 
$\F_q$. Denote its associated adelic ring by $\A$ and set 
$\O=\{a\in \A:\, a_p\in \O_p\ \forall p\}$, where $\O_p$ denotes the integer ring of the 
local field $F_p$ of $F$ at $p\in X$. 

Recall that, for an element $g\in\GL_r(\A)$, in \cite{W} (see also \cite{Su}), we 
introduce the subspace $\A^r(g)$ of $\A^r$ by
\be
\A^r(g):=\Big\{ a\in \A^r:\, ga\in \O^n\Big\},
\ee
and make the following
\begin{defn} Let $g\in \GL_r(\A)$.
\begin{enumerate}
\item [(1)] The {\it 0-th cohomology of $g$ over $F$} is defined by
\be
H^0(F,g):= \A^r(g)\cap F^r.
\ee
\item[(2)] The {\it 1st cohomology of $g$ over $F$} is defined by
\be
H^1(F,g):= \A^r/(\A^r(g)+ F^r).
\ee
Here $F^r$ is viewed as a subspace of $\A^r$ through the natural diagonal embedding 
$F^r\hookrightarrow \A^r, \ f\mapsto (f)$.
\end{enumerate}
\end{defn}

Let $\cE(g)$ be the rank $r$ locally free sheaf on $X$ associated to $g$.\footnote{For 
the detailed construction, please refer to Appendix A.} Then using the ind-pro topology 
on $\A^r$, or better on $\A$ defined by $\dis{\A=\lim_{\substack{\rightarrow\\ D}}
\lim_{\substack{\leftarrow\\ D\geq D'}}\A(D)/\A(D')}$, we have the following

\begin{thm} (\cite{W} and \cite{Su})  Let $g\in \GL_r(\A)$, and denote by $\cE(g)$ the 
associated rank $r$ locally free sheaf on $X$.
\begin{enumerate}
\item[(0)] The $\F_q$-linear space $H^0(F,g)$ and $H^1(F,g)$ are finite 
dimensional. For later use, set $h^i(F,g):=\dim^{~}_{\,\F_q}H^i(X,g)$ for $i=0,1$.
\item [(1)] There are natural $\F_q$-linear isomorphisms
\be H^0(F,g)\simeq H^0(X,\cE(g))\qqan H^1(F,g)\simeq H^1(X,\cE(g));
\ee
\item[(2)] Induced by the  natural non-degenerate  pairing (with respect to a non-trivial 
rational differential $\om_0$ on $X$)
\be
\bm
\lan\cdot,\cdot\ran_{\om_0}:&\A^r\times \A^r&\lra&\F_q\\[0.5em]
&((a_i),(b_i))&\longmapsto&\sum_{i=1}^r\om_0(a_ib_i),
\em
\ee
we have
\begin{enumerate}
\item [(a)] $\widehat{\A^r}\simeq \A^r$, where $\widehat{\A^r}$ denotes the Pontryagin 
dual group of $\A^r$ (with respect to the above ind-pro topology. 
\item[(b)] $(F^r)^\perp=F^r$;
\item[(c)] $\A^r(g)^\perp=\A^r(\iota_{(\omega_0)}\cdot g^{-1})$, where 
$\iota_{(\om_0)}\in \I:=\GL_1(\A)$ denotes an idelic element corresponding to the 
invertible sheaf $\O_X((\om_0))$, or better, to the divisor $D$.
\item[(d)] (Topology Duality)  Induced by the pairing $\lan\cdot,\cdot\ran_{\om_0}$, there 
exists a corresponding duality isomorphism
\be
\Hom_{\F_q}\left(\A^r/(\A^r(g)+ F^r),\F_q\right)\simeq 
\A^r(\iota_{(\omega_0)}\cdot g^{-1})\cap F^r.
\ee
\item[(e)] (Adelic  Riemann-Roch Theorem) 
\be
h^0(F,g)-h^1(F,g)=:\chi(F,g)=\deg(g)-r(g-1).
\ee 
Here as usual $\deg(g):=\deg(\det(g))$.\footnote{Certainly, 
$\deg(g)=\deg(\cE(g))$ and, by definition, $\deg(\det\cE(g))= \deg(\cE(\det(g))).$}
\end{enumerate}
\end{enumerate}
\end{thm}

Obviously, (d) is simply the Serre duality 
\be
H^1(X,\cE(g))\simeq H^0(X,\cE(g^{-1})((\om_0))^\vee
\ee
and (e) is nothing but the Riemann-Roch theorem
\be
\chi(X,\cE(g))=\deg(\cE(g))-r(g-1).
\ee

This theorem was first outlined by the author as a by-product of an adelic cohomology 
theory for arithmetic curves in a paper on \lq Geometry of Numbers' and now  Chapter 2 
of \cite{W}. A detailed proof can be found in  Sugahara's  thesis \cite{Su}. By saying so, 
we also should mention that most of the theorem can be proved using Chevalley's 
preparatitions in any standard text books on adeles for function fields, say,  \cite{M}, 
\cite{R}, \cite{S}.

\subsection{Rank $r$ Algebraic Geometry Codes}
Let $X$ be an integral regular projective curve on $\F_q$. Denote by $F/\F_q$ its 
field of rational functions and $\A$ the adelic ring associated to $F$. Let 
$D=\sum_{i=1}^n p_i$ be a degree $n$ divisor on $X$. This implies, in particular, that 
$p_i, \, i=1,\ldots,n$ are mutually distinct $\F_q$-rational points  of $X$. Fix a $D$-
special rational differential $\om_0$ on $X$, namely, $\om_0\in \Om_F$ is a non-trivial 
rationals differential on $X$ such that 
\be
\ord_{p_i}(\om_0)=-1\qqan \res_{p_i}(\om_0)=1\qquad\forall i=1,\ldots,n.
\ee

\medskip
Let $g=(g_p)\in \GL_r(\A)$. Then $g_{p_i}\O_{p_i}^r\subset F_{p_i}^r$ is a full rank 
$\O_{p_i}$-lattice. Since $\O_{p_i}$ is a PID, there exists $n_{ij}\in Z$, $i=1,\ldots,r$, 
such that  
\be
n_{i1}\leq n_{i2}\leq\ldots\leq n_{ir}
\qqan
g_{p_i}\O_{p_i}^r\simeq \mathrm{diag}\big(\pi_{p_i}^{n_{i1}},\ldots,\pi_{p_i}^{n_{ir}}\big)
\,\O_{p_i}^r.
\ee
Here $\mathrm{diag}(a_1,\ldots,a_r)$ denotes the diagonal matrix with diagonal 
components $a_1,\ldots, a_r$. In other words, there exists $M_i,N_i\in \GL_r(\O_{p_i})$ 
such that 
\be\label{eqag0}
Mg_{p_i}N=\mathrm{diag}(\pi_{p_i}^{n_{i1}},\ldots,\pi_{p_i}^{n_{ir}}).
\ee
It is well known that $(n_{i1},\ldots n_{ir})$ depends only on $g$ and does not depend 
on the choices of the local parameter $\pi_{p_i}$ used. For our use, we call 
$(n_{i1},\ldots n_{ir})$ the $p_i$-{\it multiple orders}. Easily,
\be\label{eqag1}
\sum_{j=1}^rn_{ij}=\ord_{p_i}\det(g).
\ee

\begin{con}\normalfont  A element $g\in \GL_r(\A)$ is called $D$-{\it balanced} if its 
$p_i$-(multiple) orders satisfy the conditions
\be
(n_{i1},\ldots n_{ir})=( 0,\ldots,0)\qquad\forall i=1,\ldots,n.
\ee
\end{con}

\begin{lem}
Let $g\in \GL_r(\A)$. 
\begin{enumerate}
\item [(1)] $g$ is $D$-special if and only if $g_{p_i}\in\GL_r(\O_{p_i})$ for all 
$i=1,\ldots,n$. Here $g=(g_{kj})_{k,j}=((g_{kj,p})_p)$.
\item [(2)] If $g$ is $D$-special, then $\big((f_j(p_1)),\ldots,(f_j(p_n))\big) $ makes 
sense for each element $f=(f_j)\in H^0(F,g)$.
\end{enumerate}
\end{lem}
\bp
(1) If $g_p\in \GL_r(\O_{p_i})$, then $g_{kj,p}\in \O_{p_i}$. This implies that 
$n_{ji}\geq 0$ for all $j$. On the other hand, since $\det (g_p)\in \O_{p_i}^*$,  
$\ord_{p_i}(\det(g))=0$. Hence, by \eqref{eqag1}, $n_{ji}\geq 0$ for all ($i$ and) $j$.

Conversely, if $(n_{i1},\ldots n_{ir})=( 0,\ldots,0)$, by \eqref{eqag0}, we have
$\mathrm{diag}(\pi_{p_i}^{n_{i1}},\ldots,\pi_{p_i}^{n_{ir}})$ and hence also $g_{p_i}$ 
belong to $\GL_r(\O_{p_i})$.

(2) Since $g_{p_i}\in \GL_r(\O_{p_i})$, $\det(g_{p_i})\in \O_{p_i}^*$, and hence
$g_{p_i}^{-1}$ belong to $\GL_r(\O_{p_i})$ as well. This implies that 
$g_{p_i}(p_i)\in\GL_r(\F_q)$. Therefore, note only
\be
(g_{p_1}(f_j)(p_1),\ldots,g_{p_n}(f_j)(p_n))
=(g_{p_1}(p_i)\,(f_j(p_1)),\ldots,g_{p_n}(p_n)\,(f_j(p_n)))
\ee
is well-defined, but the morphism
\be
\bm
g_{p_i}:&\F_q^{rn}&\lra&\F_q^{rn}\\
&(g(p_i) (f_j(p_1)),\ldots,g(p_n)(f_j(p_n))&\longmapsto&((f_j(p_1)),\ldots,(f_j(p_n)))
\em
\ee
makes sense.
\ep

Now we are ready to introduce the first main definition.

\begin{defn}\normalfont  Let $g\in \GL_r(\A)$ be  $D$-special. The {\it rank $r$ 
algebraic geometry codes associated to $g$ with respect to $D$} is defined by the 
codewords space
\be
C_{F,r}(D,g)
:=\Big\{\big((f_j(p_1)),\ldots,(f_j(p_n))\big)\in \F_q^{rn}\,|\, (f_j)\in H^0(F,g)\Big\}.
\ee
\end{defn}

Obviously, the length of the codewords in $C_r(F,g(-D))$ is $rn$. To understand these 
types of new codes, we consider the evaluation morphism
\be
\bm
\ev_{D,g}:&H^0(F,g)&\lra&C_{F,r}(D,g)\\[0.4em]
&(f_j)&\longmapsto&\big((f_j(p_1)),\ldots,(f_j(p_n))\big).
\em
\ee
By definition, $\ev_{D,g}$ is surjective. To determine its kernel $\Ker( \ev_{D,g}),$ we 
first note that for $(f_j)\in H^0(F,g)$, 
\be
\big((f_j(p_1)),\ldots,(f_j(p_n))\big)
=((0),\ldots,(0))\Llra g(f_j)\in \pi_{p_i}\O_{p_i}^r\qquad\forall i=1,\ldots,n. 
\ee
Set now $\iota_D=(\iota_{D_p})_p\in I$ be the idle associated to the divisor defined by
\be
\iota_{D_p}=\bc \pi_{p_i}&p=p_i\ (i=1,\ldots,n)\\
1&p\not\in\{p_1,\ldots,p_n\} 
\ec
\ee
In other words, $\iota_D=(\pi_p^{\ord_p(D)})_p\in \I=\GL_1(\A)$, which clearly also  
characterizes the divisor $D$. Then,  by definition of $H^0(F,\cdot)$ again,   we have
\bea
\Ker( \ev_{D,g})=&\l\{(f_j)\in F^r: g_{p_i}(f_j)\in \pi_{p_i}\O_{p_i}^r\ \forall i=1,\ldots,n\r\}\\
=&\l\{(f_j)\in F^r: g(f_j)\in \iota_{D}\O^r\r\}\\
=&\l\{(f_j)\in F^r: \iota_{D}^{-1}g(f_j)\in \O^r\r\}\\
=&\A^r(\iota_{D}^{-1}g)\cap F^r\\
=&H^0(F, \iota_{D}^{-1}g)
\eea
By an abuse of  notation, in the sequel, we denote $\iota_{D}^{-1}g$ simply by $g(-D)$. 
Then what we have just said proves the following
\begin{thm}\label{mthm1} Let $g\in \GL_r(\A)$ be $D$-special. There exists a canonical 
short exact 
sequence
\be
0\to H^0(F,g(-D))\lra H^0(F,g)\buildrel{\ev_{D,g}}\over\lra C_{F,r}(D,g)\to 0,
\ee
In particular, the dimension of $C_{F,r}(D,g)$ is equal to
\be
k_{D,g}=h^0(F,g)-h^0(F,g(-D)).
\ee
\end{thm}
\begin{ex} Let $E$ be an effective divisor on $X$ such that the characterizing idele 
$\iota_E:=(\pi_p^{\ord_p(E)})\in \GL_1(A)$  is $D$-balanced. Obviously, this latest 
condition is equivalent to the condition that  the supports of $D$ and $E$ are mutually 
disjoint. Moreover, it is not too difficult to see that
\be
C_{F,1}(D, \iota_E)=C_L(D,E).
\ee
\end{ex}

\section{Rank $r$ Differential Codes}

Before we introduce the codewords space which is dual to the rank $r$ algebraic 
geometry codes $C_{F,r}(D,g)$, we reexamine how $C_\Omega(D,E)$ is introduced.
In that case, we use the space $\Omega_F(E-D)$ which is defined by
\be
\Omega_F(E-D):=\big\{\om\in \Omega_F\,|\, (\om)\geq E-D\big\}
\ee
To go further, we fix  an $F$-linear isomorphism
\be
\bm
\phi:&\Omega_F&\lra&F\\
&\om&\longmapsto&\frac{\om}{\om_0}.
\em
\ee 
In terms of $\phi$, we have
\bea
\Omega_F(E-D)
=&\big\{\phi(\om)\cdot \om_0\in \Omega_F\,|\, \phi(\om)+(\om_0)\geq E-D\big\}\\
=&\big\{f\, \om_0\in \Omega_F\,|\, f+(\om_0)+D- E\geq 0\big\}\\
=&\big\{f\, \om_0\in \Omega_F\,|\, \iota_{(\om_0)+D}\,f\in \iota_E\,\O\big\}.
\eea

Motivated by this, we introduce the following
\begin{defn}\normalfont
The {\it rank $r$ rational differential space $\Omega_F^r(g(-D))$}
is defined by
\bea
\Omega_F^r(g(-D))
:=&
\Big\{ (\om_j)\in \Omega_F^r\,|\, \iota_{(\om_0)+D}\,(\phi(\om_j))\in g\,\O^r\Big\}\\[0.4em]
=&\Big\{ \om_0\,(f_j)\in \Omega_F^r\,|\, \iota_{(\om_0)+D}\,(f_j)\in g\,\O^r\Big\}.
\eea
\end{defn}
In particular, if $g$ is $D$-balanced, since $\ord_{p_i}(\om_0\iota_D)=0$ and $\om_0$ is
$D$-special, we certainly get 
\be\label{eqag4}
\phi(\om_j)(p_i)=\res_{p_i}(\om_0\phi(\om_j))=\om_{j,P_i}(1).
\ee
With this, we are ready to introduce the next main definition.

\begin{defn}\normalfont Let $g\in\GL_r(\A)$ be  $D$-balanced.  The {\it rank $r$ 
differential codewords space associated to $D$ and $g$} is defined by
\be
C_{\Omega,r}(D,g):=\Big\{\big((\om_{j,p_1}(1)),\ldots, (\om_{j,p_n}(1))\big)\,\big|\, 
(\om_j)\in \Omega_{F}^r(g(-D))\Big\}.
\ee
\end{defn}

To see the dimension of this codewords space, we introduce the following morphism
\be
\bm
\ev_{D,g}^\perp:& \Omega_{F}^r(g(-D))&\lra&C_{\Omega,r}(D,g)\\
&(\om_j)&\lmt&\big((\om_{j,p_1}(1)),\ldots, (\om_{j,p_n}(1))\big).
\em
\ee
Directly from the definition, $\ev_{D,g}^\perp$ is surjective. Hence to obtain the 
dimension of the codewords space, it suffices to determine its kernel 
$\Ker(\ev_{D,g}^\perp)$. Since $(\om_j)\in \Ker(\ev_{D,g}^\perp)$ if and only if 
$(\om_{j,p_1}(1))=0$ for all $i=1,\ldots n$, which, by \eqref{eqag4}, is equivalent to the 
condition that $\big(\phi(\om_j)(p_i)\big)=(0,\ldots,0)$. Therefore, 
\bea
 \Ker(\ev_{D,g}^\perp)
 =&\Big\{ (\om_j)\in \Omega_F^r\,|\, \iota_{(\om_0)+D}\,(\phi(\om_j))\in 
 \iota_D\,g\,\O^r\Big\}\\
 =&\Omega_F^r(g(-D+D))\\
 =&\Omega_F^r(g)).
\eea
This then proves the first part of the following
\begin{thm}\label{mthm2} Let $g\in \GL_r(\A)$ be $D$-balanced. 
\begin{enumerate}
\item [(1)] There is a short exact sequence of $\F_q$-linear spaces
\be
0\to \Omega_F^r(g)\lra \Omega_F^r(g(-D))\buildrel{\ev_{D,g}^\perp}\over\lra 
C_{\Omega,r}(D,g)\to 0.
\ee
\item[(2)] The dimension $k_{D,g}^\perp$ of the codes $C_{\Omega,r}(D,g)$ is equal to
\be
h^1(F,g(-D))-h^1(F,g).
\ee
\item[(3)] The dimensions of $C_{F,r}(D,g)$ and $C_{\Omega,r}(D,g)$ satisfy
\be
k_{D,g}+k_{D,g}^\perp=n\,r.
\ee
\end{enumerate}
\end{thm}
\bp
What left is the proof of (2) and (3). Assume that (2) holds. Then by Theorem\,
\ref{mthm1}, we have, by the Riemann-Roch theorem,
\bea
k_{D,g}+k_{D,g}^\perp=&\l(h^0(F,g)-h^0(F,g(-D))\r)+\l(h^1(F,g(-D))-h^1(F,g)\r)\\
=&\l(h^0(F,g)-h^1(F,g)\r)-\l(h^0(F,g(-D))-h^1(F,g(-D))\r)\\
=&\big(\deg(g)-r(g-1)\big)-\big(\deg(g)-r(g-1)\big)\\
=&\deg(g)-\deg(g(-D))\\
=&\,n\,r
\eea
since, we have, by definition,  
\bea
\deg(g(-D))=&\deg\big(\det(g(-D))\big)=\deg\big(\det(\iota_D^{-1}g)\big)\\
=&\deg\big(\det(g)\cdot \iota_D^{-r}\big)\\
=&\deg\big(\det(g)\big)+\deg\big(\iota_D^{-r}\big)\\
=&\deg\big(\det(g)\big)-r\deg\big(\iota_D\big)\\
=&\deg(g)-r\deg(D).
\eea
This proves (3) assuming (2).  

Finally, we prove (2). For this, we use the 
isomorphism obtained from the duality theorem
\be
\Omega_F^r(g)\simeq 
\Hom^{~}_{\F_q}\l(\A^r\big/\big(\A^r(g(-D))+F^r\big),\F_q\r)=H^1(F,g)^\vee.
\ee
This then completes the proof.
\ep

\begin{ex} Let $E$ be an effective divisor on $X$ such that the characterizing idele 
$\iota_E:=(\pi_p^{\ord_p(E)})\in \GL_1(A)$  is $D$-balanced, i.e.  the supports of $E$ 
and $D$ are mutually disjoint. Then, easily from the definitions of both sides below,
\be
C_{\Omega,1}(D, \iota_E)=C_\Omega(D,E).
\ee
\end{ex}

\section{Duality between AG  and Differential Codes in Rank $r$}
Theorem\,ref{mthm2}(3) suggests that, similar to canonical AG and differential codes, 
there is also a natural duality between rank $r$ AG and differential codes. This is indeed 
the case, as to be exposed in this section.

We start with the non-degenerate 
pairing 
\be
\bm
\lan\cdot,\cdot\ran:&(\F_q^r)^n\times (\F_q^r)^n&\lra&\F_q\\[0.5em]
&\big((a_{ji})_j)_i,((b_{ji})_j)_i\big)&\longmapsto&\sum_{i,j=1}^{n,r}a_{ji}b_{ji}.
\em
\ee
Obviously, this $\lan\cdot,\cdot\ran$ induces a natural pairing between subspaces 
$C_{F,r}(D,g)$ and $C_{\Omega,r}(D,g)$ of $(\F_q^r)^n$ as follows: 
\be
\bm
\lan\cdot,\ran_{D,g}:& C_{F,r}(D,g)\times C_{\Omega,r}(D,g)&\lra&\F_q\\[0.5em]
&\Big(\big((f_j(p_i))_j\big)_i,\big(\big(\om_{j,p_i}(1)\big)_j\big)_i\Big)&\longmapsto
&\dis{\sum_{i,j=1}^{n,r}}f_j(p_i)\cdot \om_{j,p_i}(1).
\em
\ee

\begin{thm} Let $g\in\GL_r(\A)$ be $D$-balanced.
\begin{enumerate}
\item [(1)] The natural pairing $\lan\cdot,\ran_{D, g}$ degenerates completely. That is,
the image of $\lan\cdot,\ran_{D, g}$ consists of only the zero element.
\item [(2)] $C_{\Omega,r}(D,g)$ is the dual codes of $C_{F,r}(D,g)$. That is to say,
\be
C_{\Omega,r}(D,g)^\perp=C_{F,r}(D,g).
\ee
\end{enumerate}
\end{thm}
\bp
Since (1) and (2) are equivalent, we only need to prove (2). 
By Theorem\,\ref{mthm2}(3), we have
\be
\dim^{~}_{\F_q}C_{F,r}(D,g)+\dim^{~}_{\F_q}C_{\Omega,r}(D,g)=n\,r.
\ee
As a direct consequence, it suffices to show that
\be\label{claim1}
C_{\Omega,r}(D,g)^\perp\supseteq C_{F,r}(D,g).
\ee
Let $\big((\om_{j,p_1})(1),\ldots, (\om_{j,p_n})(1)\big)$ be an element of $C_{F,r}(D,g)$ 
with  $(\om_j)\in \Omega^r_F(g(-D)).$ Since 
$\Omega^r_F(g(-D))=\Hom^{~}_{\F_q}\l(\A^r\big/\big(\A^r(g(-D))+F^r\big),\F_q\r)$, 
we have
\be\label{eq6.6}
(\om_j)\big(\A^r(g(-D))\big)=\{0\}\qqan (\om_j)\big(F^r\big)=\{0\}.
\ee
The second equality is equivalent to the residue formula. Indeed, if we write 
$\om_j=\om_0\cdot\phi(\om_j)$, then, for any $(f_j)\in F^r$,
\be
(\om_j)(f_j)=\sum_{j=1}^n\om_j(f_j)
=\sum_{j=1}^n\om_0(f_j\phi(\om_j))=\om_0\Big(\sum_{j=1}^nf_j\phi(\om_j)\Big)=0,
\ee
by the residue formula. So we must deduce \eqref{claim1} from the first relation in 
\eqref{eq6.6}. Take then an element $(a_j)\in \A^r(g(-D))$. By 
definition, $(a_j)\in \iota_D^{-1} g(a_j)\in \O^r$. This means that
\be\label{eq6.6.1}
\pi_{p_i}(a_{j,p_i})\in g_{p_i}^{-1}\O^r \qqan (a_{j,q}) g_q\in\O_q^r\ \ \forall q\not=p_i,\ 
\forall i=1,\ldots,n,
\ee
since $g$ is $D$-balanced. In other words, for any $(a_j)\in \A^r$  satisfying 
\eqref{eq6.6.1}, we have $(\om_j)(a_j)=0$. To go further, we take an element
$\big((f_j(p_1)),\ldots,(f_j(p_n))\big)\in C_{\Omega,r}(D,g)$ for a certain 
$(f_j)\in H^0(F,g)$. Then, by \eqref{eqag4}
\bea
&\big\lan \big((f_j(p_1)),\ldots,(f_j(p_n))\big), 
\big((\om_{j,p_1})(1),\ldots, (\om_{j,p_n})(1)\big)\big\ran\\
&=\sum_{i,j=1}^{n,r}f_j(p_i)\om_{j,p_i})(1)=\sum_{i,j=1}^{n,r}f_j(p_i)\phi(\om_j)(p_i)\\
&=\sum_{i,j=1}^{n,r}(f_j\phi(\om_j))(p_i)
=\sum_{i,j=1}^{n,r}\Big(f_j\frac{\om_j}{\om_0}\Big)(p_i).
\eea
But $\om_0$ is $D$-special means that 
\be
\ord_{p_i}(\om_0)=-1 \qqan \res_{P_i}(\om_0)=1\qquad\forall i=1,\ldots,n.
\ee
This implies that
\be
\Big(f_j\frac{\om_j}{\om_0}\Big)(p_i)=\om_{j,p_i}\Big(\frac{f_j}{\pi_{p_i}}\big).
\ee
Therefore,
\bea\label{eq6.6.4}
&\big\lan \big((f_j(p_1)),\ldots,(f_j(p_n))\big), 
\big((\om_{j,p_1})(1),\ldots, (\om_{j,p_n})(1)\big)\big\ran\\
&=\sum_{i,j=1}^{n,r}\om_{j,p_i}\Big(\frac{f_j}{\pi_{p_i}}\Big)
=\sum_{i=1}^n(\om_{j,p_i})\Big(\frac{f_j}{\pi_{p_i}}\Big)\\
&=\sum_{\substack{q:\,q\not=p_i\\ i=1,\ldots,n}}\om_{j,q}(f_{j}(q))
+\sum_{i=1}^n(\om_{j,p_i})\Big(\frac{f_j}{\pi_{p_i}}\Big)\\
&=(\om_j)(\iota_D^{-1}(f_j)).
\eea
Here, in the equality above the last, we have used the fact that
for $(\om_j)\in \Omega_{F,r}(g(-D))$ and $(f_j)\in H^0(F, g)$, 
$\om_{j,q}(f_{j}(q))=0$. Thus, by using \eqref{eq6.6.1}, we get 
$\iota_D^{-1}(f_j)\in \A^r(g(-D))$. Therefore, by  the fact that $(\om_j)(\A(g(-D)))=\{0\}$
for $(\om_j)\in \Omega_{F}^r(g(-D))$, we have $(\om_j)(\iota_D^{-1}(f_j))$. This means 
that, for all elements $(\om_j)\in \Omega_{F}^r(g(-D))$ and $(f_j)\in \A^r(g)$,
\be
\big\lan \big((f_j(p_1)),\ldots,(f_j(p_n))\big), 
\big((\om_{j,p_1})(1),\ldots, (\om_{j,p_n})(1)\big)\big\ran=0
\ee
by \eqref{eq6.6.4}. This establishes \eqref{claim1} and hence completes the proof of the 
theorem.
\ep
As a direct consequence of the proof of the theorem above, we have the following
\begin{cor} Let $g\in\GL_r(\A)$ be $D$-balanced. Then
\be
C_{\Omega,r}(D,g)=C_{F,r}(D,\iota_{(\om_0)+D}\,g^{-1}).
\ee 
where $(\om_0)$ is a $D$-special rational differential.
\end{cor}

\section{Estimate Dimensions of Rank $r$ AG Codes}
It is proved in Theorems\,\,\ref{mthm1} and \ref{mthm2}, we have determined the 
dimensions $k_{D, g}$ and $k_{D, g}^\perp$ for the rank $r$ algebraic geometry codes 
$C_{F, r}(D, g)$ and $C_{\Omega,r}(D, g)$, respectively. Namely, for $D$-balanced 
$g\in \GL_r(\A)$,
\bea
k_{D, g}=&h^0(F,g)-h^0(F, g(-D)),\\
k_{D, g}^\perp=&h^1(F, g(-D))-h^1(F,g).
\eea
As for classical AG codes, this is not good enough. Indeed, there, with a conditions that 
$\deg(E)<\deg(D)$ and $\deg(E)>2g-2$, by the vanishing results
\be
h^0(F, \iota_{E-D})=0\qqan h^1(X,\iota_E)=0,
\ee
we obtain 
\be 
k_{D, \iota_E}=h^0(F,\iota_E)\qqan k_{D, \iota_E}^\perp=h^1(F, \iota_{E-D}),
\ee
respectively, provided that the supports of $D$ and $E$ are disjoint.

By contrast, these is no such vanishing result for cohomologies of rank $r$ settings.

\begin{ex}\normalfont Let $g=\diag(\iota_{D_1}, \iota_{D_2})$. Then
\be
h^i(F, g)=h^i(F,\iota_{D_1})+h^i(F,\iota_{D_2})\qquad i=0,1.
\ee
There is {\it no simple yet general vanishing result} for $h^i(F,g)$ depending only on the 
degree $d$ of $g$, since the degrees of the divisors $D_1$ and $D_2$ can be changed 
freely as long as they satisfy $\deg(D_1)+\deg(D_2)=d$. Hence, even 
$h^i(F,\iota_{D_j})$ admit vanishing properties, but not $h^i(F,g)$.
\end{ex}

This is not the end of our theory. Much better, in algebraic geometry, there is a powerful 
and general vanishing theory based on the stability.
\begin{defn} 
\begin{enumerate}
\item [(1)] {\bf (Mumford \cite{Mu})} \normalfont A locally free sheaf $\cE$ on $X$ is 
called {\it semi-stable}, resp. {\it stable}, if for any proper subsheaf $\cF$ of $\cE$, 
\be
\mu(\cF)\leq \mu(\cE)\quad rasp.\quad \mu(\cF)< \mu(\cE)
\ee
Here as usual, the $\mu$-slope  is defined by
\be
\mu(\cF):=\frac{\deg(\cF)}{\rk (\cF)}.
\ee
\item [(2)] An element $g\in \GL(\A)$ is called {\it semi-stable}, resp. {\it stable}, if its 
associated locally free sheaf $\cE(g)$ is  semi-stable, resp.  stable.
\end{enumerate}
\end{defn}
It is not too difficult to prove the following:
\begin{lem} Let $\cE$ be a semi-stable local free sheaf on $X$ of rank $r$. 
\begin{enumerate}
\item [(1)] If $\deg(\cE)>r(2g-2)$, then $H^1(X,\cE)=\{0\}$.
\item [(2)] If $\deg(\cE)<0$, then $H^0(X,\cE)=\{0\}$.
\end{enumerate}
\end{lem}
\bp
Indeed, if $\deg(\cE)<0$ and $H^0(X,\cE)\not=\{0\}$, there exists a non-trivial global 
section $s$ of $\cE$. Hence, we obtain an injection $\O_X\buildrel{s}\over\lhra\cE$.
By the semi-stability condition on $\cE$,
\be
\deg(\O_X)\leq \frac{\deg(\cE)}{r}.
\ee
In particular, $\deg(\cE)\geq 0$. This proves (2) and hence also (1) by the duality 
theorem.
\ep
As a direct consequence, using the duality and the Riemann-Roch theorem, we obtain 
the following
\begin{cor}\label{cor6.8} Let $g\in \GL_r(\A)$ be $D$-balanced. Assume that $g$ is 
semi-stable. 
\begin{enumerate}
\item [(1)] If $\deg(g(-D))<0$, namely, $\deg(g)<r\deg(D)$, then
\be
k_{D, g}=h^0(F,g)\geq \deg(g)-r(g-1).
\ee
\item [(2)] If $\deg(g)>2r(g-1)$ then
\be
k_{D, g}^\perp=h^1(F,g(-D))\geq r\big(\deg(D)+(g-1)\big)-\deg{g}.
\ee
\item [(3)] If $r\deg(D)>\deg(g)>2r(g-1)$ then
\be
k_{D, g}= \deg(g)-r(g-1)\qqan k_{D, g}^\perp= r\big(\deg(D)+(g-1)\big)-\deg{g}.
\ee
\end{enumerate}
\end{cor}

\section{Masses of Semi-Stable Locally Free Sheaves}
To have a rich theory for rank $r$ algebraic geometry codes, there is a problem to find 
how many $\F_q$-rational semi-stable locally free sheaves of rank $r$ and degree $d$.
To answer this, (not really as usual), we denote by $\cM_{X,r}(d)$ the moduli 
stack of $\F_q$-rational semi-stable locally free sheaves of rank $r$ and degree~$d$.
\begin{defn}[\cite{HN}]\normalfont For each pair $(n,d)$, we define 
the {\it $\beta$-invariant} $\beta(r,d)$  by
\be
\beta_{r,d}:=\sum_{[\cE]\in \cM_{X,r}(d)}\frac{1}{|\Aut(\cE)|}.
\ee
\end{defn}
Then, staring from the fact that the Tamagawa number of $\SL_r(\A)$ is one, using 
parabolic reduction, Harder-Narasimhan can calculate $\beta_{n,d}$ for all $d$. For 
example, with additional works of Desale-Ramanan and Zagier, we have the 
following well-known formula for $\beta_{r,0}$.  
\begin{thm}[\cite{HN}, see also \cite{W0} ] For any integer $\a$,
\be
\beta_{r,r\a}=\sum_{k=1}^r(-1)^k
\sum_{\substack{n_1,\ldots, n_k\in \Z_{>0}\\ n_1+\ldots+ n_k=r}}
\frac{\prod_{i=1}^k \widehat v_{X,n_i}}{\prod_{j=1}^{k-1}q^{n_j+n_{j+1}}-1}.
\ee
Here $ \widehat v_{X,n}:=\whz_X(n)$ with $\whz_X(s)=q^{(g-1)s}\zeta_X(s)$ the 
(complete) Artin zeta function for $X$.\footnote{We here set 
$\whz_X(1)=\res^{~}_{s=1}\whz^{~}_X(s)$.}
\end{thm}

Back to elements $g\in \GL_r(A)$, we mention that Lafforgue obtains an analytic 
characterization for $g$ to be semi-stable using Arthur's analytic truncation, a 
fundamental tool introduced by Arthur in his study of trace formula. For details, please 
refer to \S V.1 of \cite{L}. This result is further generalized by the author to general 
reductive groups (\cite{W}).

\section{Minimal Distances of Rank $r$ AG Codes}

To begin with, we first recall the following well-known
\begin{defn} Let $C$ be a linear code and let $a,\,b$ two codewords  of $C$. 
\begin{enumerate}
\item [(1)] The (Hamming) distance $d(a,b)$ between $a$ and $b$  is the minimum 
number of coordinate positions in which they differ.
\item [(2)]  The (Hamming) weight $w(a)$ of  $a$ is the number of coordinate positions 
which are non-zero.
\item [(3)] The minimal distance $d(C)$ of $C$ is the weight of the smallest weight non-
zero codewords. That is,
\be
d(C)=\min_{a,b\in C, a\not=b}d(a,b)=\min_{a\in C, a\not=0}w(a).
\ee 
\end{enumerate}
\end{defn}

Our main aim in this section is to study the minimal distance of the rank $r$ algebraic 
geometry codes $C_{F, r}(D, g)$ and hence also for $C_{\Om,r} (D, g)$ for $D$ 
balanced element $g\in \in \GL_r(\A)$.

\begin{lem}
Let $f=(f_j)\in H^0(F,g)$. Then the weight of the codeword 
$\big((f_j(p_1)),\ldots (f_j(p_n))\big)\in \CFr$ is given by
\be
w\big((f_j(p_1)),\ldots (f_j(p_n))\big)
=nr-\sum_{i,j=1}^{n, r}\delta^{~}_{\ord_{p_i}(f_j)\geq 1}.
\ee
where, for $a,b\in \R$, $\delta_{a\geq b}$ is defined to be 0 if $a<b$ and 1 if $a\geq b$
\end{lem}
\bp
This comes  directly from the definition of the weight of a codeword, since 
\bes
f_j(p_i)=0\qquad \mathrm{if\ and\ only\ if}\qquad \delta^{~}_{\ord_{p_i}(f_j)\geq 1}.
\ees
\vskip -0.70cm
\ep

Denote by $d_{D, g}$, resp. $d_{D, g}^\perp$, be the minimal distance of $\CFr.$
Then 
\bea\label{eq680}
d_{D, g}=&\min\Big\{ nr-\sum_{i,j=1}^{n, r}\delta^{~}_{\ord_{p_i}(f_j)\geq 1}\,\big|\,
f=(f_j)\in H^0(F,g)\Big\}\\
=&\,nr-\max\Big\{\sum_{i,j=1}^{n, r}\delta^{~}_{\ord_{p_i}(f_j)\geq 1}\,\big|\,
f=(f_j)\in H^0(F,g)\Big\}.
\eea

To go further, recall that there is a natural upper bound 
\be
k_{D, g}+d_{D, g}\leq rn+1
\ee
coming from the singleton bound for linear codes. Our aim next is to obtain a general 
lower bound for $k_{D, g}+d_{D, g}$. Thus, if we assume that in addition that $g$ is 
semi-stable and $\deg(g(-D))<0$, then $k_{D, g}\geq \deg(g)-r(g-1)$ by 
Corollary\,\ref{cor6.8}. So it suffices to find a universal lower bound for 
$d_{D, g}+\deg(g)$. For this, we reexamine the condition that $f=(f_j)\in H^0(F,g)$, i.e.
\be
g_p(f_j)\in \O_p^r\qquad \forall p.
\ee
Since $\O_p$ is a PID, it makes sense for us to talk about the multiple order 
$(n_{p1},\ldots, n_{pr})$ of $g$ at $p$. That is, there exists $M_p,\, N_p\in \GL_r(\O_p)$ 
such that
\be
g_p=M_p\diag\big(\pi_p^{n_{p1}},\ldots,\pi_p^{n_{pr}}\big)N_p.
\ee
Note that here the condition $n_{p1}\leq \ldots\leq n_{pr}$ is dropped so that
\be
g_p(f_j)\in\O_p^r\qquad \mathrm{if\ and\ only\ if}\qquad  
\ord_p(N_{pk}(f_j))+n_{pk}\geq 0\quad \forall k=1,\ldots,r.
\ee
Here  $N_{pk}$ denotes the $k$-th cow if $N_p$.
This implies that
\be
\deg(g)=\sum_{p}\sum_{j=1}^r n_{pj}\deg(p)\qan 
\deg(g)+\sum_{k=1}^r\sum_{p}\ord_{p}\big(N_{pk}(f_j)\big)\deg(p)\geq 0,
\ee
as these summations involves only finitely many $p$.

Choose now $f_0=(f_{0j})\in H^0(F,g)$ such that
\be\label{eq681}
\sum_{i,j=1}^{n, r}\delta^{~}_{\ord_{p_i}(f_{0j})\geq 1}=
\max\Big\{\sum_{i,j=1}^{n, r}\delta^{~}_{\ord_{p_i}(f_j)\geq 1}\,\big|\,
f=(f_j)\in H^0(F,g)\Big\}.
\ee
Then
\be
\deg(g)+\sum_{i,j=1}^{n,r}\ord_{p_i}(f_{0j})
+\sum_{j=1}^r\sum_{p\not\in \{p_1,\ldots,p_n\}}\ord_{p}(f_{0j})\deg(p)\geq 0
\ee
In fact much better can be done. To see this, introduce the following
\begin{defn}\normalfont  Let $f=(f_j)\in H^0(F,g)$ be a global section. 
\begin{enumerate}
\item [(1)]  The {\it margin adelic element $\chi(f,D)\in \GL_r(O)$  of $f$ with respect to 
$D$} is defined by 
\bes
\chi(f,D)_p:=\bc \diag(1,\ldots,1)& p\not\in \{p_1,\ldots, p_n\}\\[0.5em]
N_p^{-1}\diag\big(\pi_p^{-\delta_{\ord_p(f_1)\geq 1}},\ldots,
\pi_p^{-\delta_{\ord_p(f_r)\geq 1}}\big)&p\in \{p_1,\ldots,p_n\}.
\ec
\ees
\item [(2)] The {\it logarithmic transform $g_{\log(D,f)}$ of $g$ with respect to $(D,f)$} is 
defined by $g\cdot \chi(f,D)$. Namely,
\bea
&g_{\log(f,D),p}\\
&:=\bc g_p&p\not\in \{p_1,\ldots, p_n\}\\[0.5em]
M_p\,\diag\big(\pi_p^{-\delta_{\ord_p(f_1)\geq 1}},\ldots,\pi_p^{-\delta_{\ord_p(f_r)
\geq 1}}\big)
&p\in \{p_1,\ldots, p_n\}
\ec
\eea 
\end{enumerate}

\end{defn}

\begin{lem} With the same notation as above, we have
\be
f_0\in H^0\big(F,g\cdot \chi(f_0,D)\big).
\ee
In particular, if $g$ is semi-stable, then $g\cdot \chi(f_0,D)$ is also semi-stable.
\end{lem}
\bp If $p\not\in\{p_1,\ldots, p_n\}$, the $p$ component $\big(g\cdot \chi(g,D)^{-1}\big)_p$ 
of $g\cdot \chi(g,D)^{-1}$ is simply $g_p$,
Hence,
\be
g\cdot \chi(g,D)^{-1}f_0\in\O_p^r.
\ee
Now assume that $p=p_i$ for a certain $i=1,\ldots,n$. Then
\bea
&\big(g\cdot \chi(g,D)^{-1}\big)_pf_0=g_{\log(f,D),p}f_0\\
&=M_p\,
\diag\big(\pi_p^{-\delta_{\ord_p(f_1)\geq 1}},\ldots,\pi_p^{-\delta_{\ord_p(f_r)\geq 1}}\big)
f_0\\
&=M_p\big(\pi_p^{-\delta_{\ord_p(f_1)\geq 1}}f_1,\ldots,\pi_p^{-\delta_{\ord_p(f_r)
\geq 1}}f_j\big)^t\in \O_p^r
\eea
since $\ord_p\big(\pi_p^{-\delta_{\ord_p(f_r)\geq 1}}f_j\big)\geq 0$ for all $j=1,\ldots,r$.
This prove that $f_0\in H^0\big(F,g\cdot \chi(g,D)\big)$. To prove the semi-stablity 
statement, we note that the correspondences between sub-sheaves of $\E(g)$ and 
$\E(g\cdot \chi(f,D))$ in terms of their stalks at each points. Hence,  if $g$ is 
semi-stable, so is $g\cdot \chi(f_0,D)$, since $f_0\in H^0\big(F,g\cdot \chi(f_0,D)\big).$
\ep
\begin{thm} Let $g\in \GL_r(\A)$ be $D$-balanced and semi-stable. Then the minimal 
distance $d_{D, g}$ of the rank $r$ algebraic geometry code has the following lower 
bound.
\be
d_{D, g}\geq nr-\deg(g)
\ee
In particular, if $\deg(g)<rn$, we have
\be
r(n-(g-1))\leq k_{D, g}+d_{D, g}\leq rn+1
\ee
\end{thm}
\bp
Since $H^0\big(F, g\cdot \chi(g,D)\big)\not= \{0\}$, we have, by \eqref{eq681} 
and \eqref{eq680},
\bea
0\leq &\deg\big(g\cdot \chi(g,D)\big)=\deg(g)+\deg(\chi(D, g))\\
=&\deg(g)-\sum_{i,j=1}^{n, r}\delta^{~}_{\ord_{p_i}(f_{0j})\geq 1}\\
=&\deg(g)-\max\Big\{\sum_{i,j=1}^{n, r}\delta^{~}_{\ord_{p_i}(f_j)\geq 1}\,\big|\,
f=(f_j)\in H^0(F,g)\Big\}\\
=&\deg(g)+d_{D, g}-nr
\eea
With this, then  the last statement is simply that of Corollary\,\ref{cor6.8}.(1). 
\ep

Certainly, if $g=\iota_E$, this result coincides with the well-known bounds for the 
classical AG codes $C_L(D,E)$ claiming that
\be
n+1-g\leq k_{D, g}+d_{D, g}\leq n+1.
\ee

\begin{appendices}

\section{\\ \hskip -3.60cm Adelic Interpretations of Locally Free Sheaves}
For each $g=(g_p)\in\GL_r(\A)$, we obtain a natural family of lattices 
$\{g^{-1}(\O_p^r)\}_{p\in X}.$ In other words, for each $p\in X$, 
$g^{-1}(\O_p^r)\subset F_p^r$ is a full rank $\O_p$-lattice in $F_p$. Here, as usual, 
$(F_p,\O_p,\frak m_p.\pi_p)$ denotes the local field $F_p$, the local ring $\O_p$ of 
integers, the maximal ideal $\frak m_p$ and  a local parameter $\pi_p$, of $X$ at $p$. 
In parallel, denote by $(F_{ p},\O_{ p},\frak m_{ p}.\pi_p)$ the corresponding 
data after taking the completion.

To obtain an adelic interpretation of locally free sheaves, for each full rank 
$\O_{ p}$-lattice $\cM_{ p}$ of $F_{ p}^r$, we introduce a skyscraper sheaf 
$\cM_{ p}$ on $X$ by 
\be
U\longmapsto \cM_{ p}(U):=\bc  \cM_{ p}& p\in U\\[0.6em] 0&p\not\in U.\ec
\ee
Accordingly, $\l\{ g_p^{-1}(\O_{ p}^r)(U)\r\}_p$ makes sense. In addition, we obtain a 
sheaf $\cE(g)$ on $X$ defined by
\be
\cE(g): \ \ U\longmapsto \cF^r(U)\cap\l(\cap_{_{p\in U}} g_p^{-1}(\O_{ p}^r)(U)\r).
\ee
Here, $\cF^r$ denotes the constant sheaf on $X$ associated to $F^r$. 
We have the following well-known:
\begin{lem}\label{al1} (see e.g. \cite{H})  $\cE(g)$ is a rank $r$ locally free sheaf on $X$.
\end{lem}
\bp By definition, $\cE(g)_{\ov P}=g_p^{-1}(\O_{ p}^r)$.  Hence by Ex. 5.7(b) of Ch.2 in 
\cite{H},  it suffices to show that 
$\cE(g)$ is a coherent  $\O_X$-sheaf coherent. This is a local problem. Choose then a 
point $p\in X$. It is not too difficult to prove that, see e.g. Lemma 6.2 of \cite{Su}, there 
exist $g_{1,p}\in\GL_r(F)$ and $g_{2,p}\in \GL_n(\O_p)$ such that 
$g_p^{-1}=g_{1,p}g_{2,p}$. Choose then an affine open neighborhood $U_P$ of $p$ 
such that for all  $q\in U, \ q\not= p$, $g_q\in \GL_r(\O_q)$ and 
$A\in \GL_r(\O_q)\subset\GL_r(\O_{\ov q})$.  This is possible since
there exists only finitely many $q\in X$ such that $g_p\not\in \GL_r(\O_{\ov q})$ and 
$A\not\in  \GL_r(\O_{\ov q})$. Consequently,
\bea
\cE(g)|_U=&\cF^r|_U\cap\l(\cap_{_{p\in U}} g_p^{-1}(\O_{ p}^r)|_U\r)
=\cF^r|_U\cap\l(\cap_{_{p\in U}} g_p^{-1}(\O_{ p}^r|_U)\r)\\
=&\cF^r|_U\cap\l(\cap_{_{p\in U}} (g_{1,p}g_{2,p})(\O_{ p}^r|_U)\r)
=\cF^r|_U\cap\l(\cap_{_{p\in U}} g_{1,p}(\O_{ p}^r|_U)\r)\\
=&g_{1,p}(\cF^r|_U)\cap\l(\cap_{_{p\in U}} g_{1,p}(\O_{ p}^r|_U)\r)
=g_{1,p}\l(\cF^r|_U)\cap\l(\cap_{_{p\in U}} \O_{ p}^r|_U\r)\r)\\
=&g_{1,p}(\O_U^r).
\eea
Therefore, $\cE(g)$ is coherent and hence locally free.
\ep
Denote by $\cM_{X,r}$ be the moduli stack of (isomorphism classes of) rank $r$ locally 
free sheaves on $X$. Then we have the following well-known
\begin{prop} There is a natural bijective correspondence 
\be
\bm \pi:& \GL_r(F)\backslash \GL_r(\A)/\GL_r(\O)&\lra &\cM_{X,r}\\[0.6em]
&[g]&\longmapsto& [\cE(g)]
\em
\ee
\end{prop}
\bp We first prove that $\pi$ is well-defined. Assume that $g,\,h\in \GL_r(\A)$ satisfy
$\E(g)=\E(h)$. Then, for each $p\in X$,  $g^{-1}_p(\O_{ p}^r)= h^{-1}_p(\O_{ p}^r)$, 
or equivalently, $(h_pg^{-1}_p)(\O_{ p}^r)= \O_{ p}^r$. This implies that 
$hg^{-1}\in \GL_r(\O)$. More generally, assume that  $\E(g)\simeq \E(h)$, this induces 
an isomorphism $\phi_\eta: \E(g)_\eta \simeq \E(h)_\eta$. Since $\E(g)_\eta\simeq F^r$ 
and $\E(h)_\eta\simeq F^r$, $\phi_K$ is determined by an element $\Phi\in \GL_r(F)$. 
Obviously, for each $p\in X$, 
\bes
\Phi\l(g^{-1}_p(\O_{ p}^r)\r)\simeq g^{-1}_p(\O_{ p}^r)\simeq h^{-1}_p(\O_{ p}^r).
\ees
Hence $\pi$ is not only well-defined, but injective.

Next we prove that $\pi$ is an surjection. Let $\E$ be a rank $r$ locally free sheaf on 
$X$. Then $\cE_{ p}\subset F_{ p}$ is a rank $r$ projective $\O_{ p}$-module.
Therefore, there exists an element $g_p\in \GL_r(F_{ p})$ such that 
$g_p(\cE_{ p})=\O_{ p}^r$. But for all but finitely many $p\in X$, 
$\cE_{ p}\simeq \O_{ p}^r$. 
This implies that, for such a $p$, $g_p\in \GL_r(\O_{ p})$. Therefore,
$g:=(g_{ p})\in \GL_r(\A)$. On the other hand, by definition, $\cE(g)\simeq \cE$.
\ep

\begin{ex}\label{ae1} \normalfont 
Let $D=\sum_pn_p\,p$ be a divisor on $X$ and denote by $\O_X(D)$ the invertible 
sheaf on $X$ associated to $D$. To give an adelic interpretation, we set 
$g_p=\pi_p^{n_p}$ and $g_D=(g_p)$. This implies that 
\be
g_p^{-1}\O_p=\pi_p^{-n_p}\O_p\simeq \O_X(D)_p.
\ee
In other words $\cE(g_D)=\O_X(D)$.
In addition,
\bea
H^0&\big(X,\cE(g_D)\big)=\big\{f\in F: gf\in \O\big\}\\
=&\big\{f\in F: g_pf\in \O_p\ \forall p\big\}
=\big\{f\in F: \pi_p^{n_p} f\in \O_p\ \forall p\big\}\\
=&\big\{f\in F: \ord_p\big(\pi_p^{n_p} f\big)\geq 0\ \forall p\big\}
=\big\{f\in F: \ord_p( f)+n_p\geq 0\ \forall p\big\}\\
=&\big\{f\in F:(f)+D\geq 0\big\}=H^0\big(X,\O_X(D)\big).
\eea
\end{ex}
This shows that it is equally easy to use adelic language, instead of locally sheaves.

\section{\\ \hskip -3.60cm Review of Classical Algebraic Geometry Codes}
As in the previous appendix, we claim no credit  but accept any possible mistakes for 
the contents here. In fact, the materials can found in \cite{M}, \cite{NX}, \cite{S} 
and \cite{TV}.
\subsection{AG codes in terms of $H^0$}\label{ae2} 
Let $D=p_1+\ldots +p_n$ be a degree $n$ divisor and let $E$ be a 
positive divisor. Assume that the $p_i$'s are mutually distinct and that  
$|E|\cap |D|=\emptyset$, where $|\cdot |$ denotes the support of the divisor.  Then by 
Example\,\ref{ae1},
\be
g_D=(g_p)\qqan g_E=(\pi_p^{\ord_p(E)})
\ee
where $g_p=\bc \pi_{p_i}& p=p_i\\
1&p\not\in \{p_1,\ldots, p_n\}\ec$. In addition,
\bea\label{aeq1}
H^0(X,\O_X(E))
=&\l\{f\in F: \bm\pi_p^{\ord_p(E)} f\in \O_p&\ \forall  p\not\in\{p_1,\ldots,p_n\},\\
\qquad\quad\, f\in \O_p&\ \forall p\in\{p_1,\ldots,p_n\}\em \r\},\\[0.6em]
H^0(X,\O_X(E-D))
=&\l\{f\in F: \bm\pi_p^{\ord_p(E)} f\in \O_p&\ \forall  p\not\in\{p_1,\ldots,p_n\},\\
\qquad\qquad \ f\in \pi_p\O_p&\ \forall p\in\{p_1,\ldots,p_n\}\em\r\},\\
\eea 
Consequently, if $f\in  H^0(X,\O_X(E))$, then $f\in \O_{p_i}$ for all $i=1,\ldots, n$. This 
implies that $(f(p_1),\ldots, f(p_n))$ makes sense. This then leads to the space
\be
C_L(D,E):=\big\{(f(p_1),\ldots, f(p_n))\in \F_q^n: f\in H^0(X,\O_X(D))\big\}.
\ee
Moreover, the natural morphism
\be
\bm \phi_{D,E}:& H^0(X,\O_X(D))&\lra &C_L(D,E)\\[0.5em]
&f&\longmapsto& (f(p_1),\ldots, f(p_n))\em
\ee
is surjective by definition, and its kernel is given by
\be
\Ker( \phi_{D,E})=H^0(X,\O_X(E-D))
\ee
by the description of $H^0(X,\O_X(E-D))$ in \eqref{aeq1}.
This gives the short exact sequence
\be\label{req-2}
0\to H^0(X,\O_X(E-D))\to H^0(X,\O_X(E))\buildrel{\phi_{D,E}}\over\lra C_L(D,E)\to 0.
\ee

This shows that the description of algebraic geometry code is equally clear in terms of  
adelic language.

\subsection{Codes in terms of $\Omega$}

Let $D$ and $E$ be the same as in Example\,\ref{ae2}. As considered in algebraic 
geometry code, we consider 
\be
C_\Omega(D,E)
:=\big\{(\om_{p_1}(1),\ldots,\om_{p_n}(1))\in \F_q^n: \ \om\in \Omega_F(E-D)\big\}.
\ee
Here $\Omega_F(E-D)$ is viewed as a collection of rational differentials $\omega$ such 
that $(\omega)\geq E-D.$ In terms of cohomology theory, we have 
\be
\Omega_F(E-D)\simeq H^0(X,K_X(D-E)).
\ee
However, with this expression, it is difficult to see how 
$(\om_{p_1}(1),\ldots,\om_{p_n}(1))$ can be defined. To explain this, we use the duality
\be
H^0(X,K_X(D-E))\simeq H^1(X,\O_X(E-D))^\vee.
\ee
Recall that $H^1(X,\O_X(E-D)):=\A/(\A(E-D)+F)$ and hence 
\be
H^1(X,\O_X(E-D))^\vee:=\Hom_{\F_q}\left(\A/(\A(E-D)+F),\F_q\r).
\ee
In this language, $\omega\in \Omega_F(E-D)$ if and only if the morphism 
$\omega:\A\to \F_q$ satisfies the condition that 
\be
\om(\A(E-D))=0\qqan \om(F)=0.
\ee
Obviously, $\omega:\A\to \F_q$ induces $\omega_p:F_{ p}\to \F_q$ by restricting 
$\omega$ to $\A_q=F_{ p}$. Moreover, for each $a\in \A$, there are only finitely many 
$p$ such that $p\not\in \O_p$, thus, for the rest almost all $p$'s,   $\omega_p(a_p)=0.$
In this way, we have
\be
\omega(a)=\sum_p\om(a_p).
\ee
Recall that the space $\Omega_F$ of all rational differentials is one dimension over $F$. 
We may and hence will write $\om=hd\pi$ for some $h\in F$.
Hence $\om_p(a_p)=\res_p(a_ph)$. Therefore, the second equation that 
$\om(F)=0$ is equivalent to
\be
\omega(f)=\sum_p\res_p(fh)=0 \quad\forall f\in F.
\ee
This is nothing but  the well-known {\it residue formula}.

To understand the first relation, we make soe preparations.
First, since $s\in H^0(X,K_X(D-E))$, we have $(s)+(\om_0)+ (D-E)\geq 0$, where 
$\omega_0$ denotes a rational section of the canonical sheaf $K_X$. For later use, set 
$(\om_0)=W_0$. This is equivalent to $(s\om_0)\geq E-D$. Since 
$\omega=s\omega_0$, this implies that 
\be\label{aeq4}
\res_p(h\pi_p^{\ord_p(D-E)+n_+})=0\qquad\forall p\in |D|\cup|E|,\ \ 
\forall n_+\in \Z_{\geq 0}.
\ee
Secondly, we go back to the definition of $\A(E-D)$: 
\bea\label{aeq3}
\A(E-D)=&\big\{a\in \A: (a)+E-D\geq 0\big\}\\
=&\l\{a\in \A:\ \bm a_p\in \O_p^*\qquad\qquad &p\not\in|E|\cup|D|\\
a_p\in\pi^{-\ord_p(E)}\O_p&p\in |E|\qquad\\
a_p\in\pi_p\O_p\qquad\quad&p\in |D|\qquad
\em \r\}.
\eea

Now, to see $\om(\A(E-D))=0$, we make the following calculation. 
\bea
&\sum_p\res_p\om( \A(E-D))\\
=&\sum_{p\not\in |E|\cup|D|}\res_p( h\O_p^*)
+\sum_{p\in |E|}\res_p(h\pi^{-\ord_p(E)}\O_p)+\sum_{p\in |D|}\res_p(h\pi_p\O_p)\\
=&\sum_{p\not\in |E|\cup|D|}\res_p( h)
+\sum_{p\in |E|}\res_p(h\pi^{-\ord_p(E)}\O_p)+\sum_{p\in |D|}\res_p(h\pi_p\O_p)\\
=&\sum_{p\not\in |E|\cup|D|}\res_p( h).
\eea
by \eqref{aeq4}. On the other hand, $(\om)\geq E-D$ implies that for 
$p\not\in |E|\cup|D|$, $\ord_p(h)\geq 0$. Therefore,
\be
\sum_p\res_p\om( \A(E-D))=\sum_{p\not\in |E|\cup|D|}\res_p( h)=0.
\ee
This proves the following well-known
\begin{cor}[Duality Theorem] With the same notation as above,
\bea
 &H^0(X,K_X(D-E))\simeq F\cap \A(W_0+D-E))\\
 &\simeq \Hom^{~}_{\F_q}(\A/(\A(E-D)+F), \F_q)\simeq H^1(X, \O_X(E-D))^\vee.
\eea
\end{cor}

Now we analysis the space $C_\Omega(D,E)$. For $\om\in \Omega_F(E-D)$, as above,
viewing it as a morphism $\A\to \F_q$, we obtain a morphism $\om_p:F_{ p}\to\F_p$.
In particular, if $p\in |D|$, we have $\om_p(\pi_p\O_p)=0$. Therefore, 
$\om(\O_p)=\om_p(\F_q)=\F_q\om_p(1).$
This implies that the map
\be
\bm
\varphi_{D,E}:&\Omega_F(E-D)&\lra&C_\Omega(D,E)\\
&\om&\longmapsto& (\om_{p_1}(1),\ldots, \om_{p_n}(1))
\em
\ee
is surjective. To see the kernel of $\varphi_{D,E}$, we need to see for which 
$\om\in \Omega_F(E-D)$, $\om_p(1)=0$ for all $p\in|D|$. For this, we use the duality 
theorem to see that
\bea
\Omega_F(E-D)=&H^1(X,\O_X(E-D))^\vee\\
=&\Hom_{\F_q}\left(\A/(\A(E-D)+F),\F_q\r)\\
=&H^0(X,K_X(D-E)).
\eea
In other words, as mentioned above, $\om\in \Omega_F(E-D)$ if and only if
the $\om$-image  is zero on both $F$ and the space 
\be
\A(E-D)=\l\{a\in \A:\ \bm a_p\in \O_p^*\qquad\qquad &p\not\in|E|\cup|D|\\
a_p\in\pi_p^{-\ord_p(E)}\O_p&p\in |E|\qquad\\
a_p\in \pi_p\O_p\quad\qquad&p\in |D|\qquad
\em \r\}.
\ee
In particular,  
\be
\om_p(\O_p)=\om_p(\F_q+\pi\O_p)=\om_p(\F_q)=\F_q\,\om_p(1),\qquad\forall p\in |D|
\ee
since $\om$ is $\F_q$-linear. With a similar discussion,
if $\om$ in the  kernel of $\varphi_{D,E}$, then not only the $\om$-image  of both $F$ 
and $\A(E-D)$ is zero, $\om_p(\O_P)=\{0\}$. That is to say, the $\om$-image  is zero on 
both $F$ and the space 
\be
\l\{a\in \A:\ \bm a_p\in \O_p^*\qquad\qquad &p\not\in|E|\cup|D|\\
a_p\in\pi_p^{-\ord_p(E)}\O_p&p\in |E|\qquad\\
a_p\in \O_p\quad\qquad&p\in |D|\qquad
\em \r\},
\ee
which is nothing but $\A(E)$. This then establish the following
\begin{thm}\label{at16} There is a short exact sequence
\be\label{req-3}
0\to \Omega_F(E)\to \Omega_F(E-D)\buildrel{\varphi_{D,E}}\over\lra C_\Omega(D,E)
\to 0.
\ee
\end{thm}

As a direct consequence of the above discussion, we also obtain the following technical 
result, which plays a central role in the  classical approaches of AG codes.
\begin{cor}\label{ac3} (Proposition 1.7.3 of \cite{S})  Let $\om\not= 0$ be a Weil 
differential of $F/\F_q$ and $p\in X$ be a closed point. Then
\be
v_p (\om) = \max\big\{ r \in \Z : \om_p(f)=0 \ \forall f\in F\ \mathrm{s.t.}\ v_p (f) 
\geq  -r \big\} .
\ee
In particular, if  $v_p (\om) \geq  -1$, then
$\om_p (1) = 0$ if and only if $v_p (\om ) \geq  0$.
\end{cor}
\bp
In fact, if we set $E=n\,p+E'$ with $|E|=\{p\}\cup|E'|$, then, by the above discussion, for 
the place $p$ concern, $\om\in \Omega(E)$ if and only if $\om_p(\pi^{-n}\O_p)=0.$
\ep

\subsection{$\Omega$ codes are AG codes}
We use the same notation as in the previous subsubsections.
By the duality theorem,  the exact sequence in Theorem\,\ref{at16} becomes the 
following
\be
0\to H^0(X,K_X(-E))\to H^0(X,K_X(D-E))\buildrel{\varphi_{D,E}^\vee}
\over\to C_\Omega(D,E)\to 0.
\ee
So the point is to see whether $(\om_{p_1}(1),\ldots,\om_{p_n}(1))$ for 
$\om\in \Omega_F(E-D)$ can be written as $(f(p_1),\ldots, f(p_n)$ for 
$f\in H^0(X,K_X(D-E))$ under the correspondence
\be
\bm
\Phi: &\Omega_F(E-D)&\lra&H^0(X,K_X(D-E))\\
&\omega&\longmapsto&\om/\om_0=:f
\em
\ee
where $\om_0$ is a suitable rational differential on $X$.
To obtain a 'proper' $\om_0$, we first see what are the properties which $\om_0$ should 
satisfies. 

Set  $H=(\om_0)+E-D$.  Then for $p\in |D|$, 
\bea\label{aeq6}
\F_q\om_p(1)=&\om_p(\F_p)=\om_p(\F_p+\pi_p\O_p)\\
=&\om_p(\O_p)=(f\om_0)_p(\O_p)
\eea
By definition, 
\be\label{aeq5}
\ord_p(f)+\ord_p(\om_0)=\ord_p(f\om_0)=\ord_p(\om)\geq -1.
\ee
In addition, $\om_p(1)=0$ if and only if $f(P)=0$. But $\om_p(1)=0$ means that as far as 
the point $p$ concern, $\om_p(\O_p)=0$, hence $\ord_p(\om_p)\geq -1$. Similarly, 
$f(p)=0$ means that $\ord_p(f)\geq 1$.  Thus, in this case, by \eqref{aeq5},
\be
0=\ord_p(f)+\ord_p(\om_0)\geq 1+\ord_p(\om_0)
\ee
That is, 
\be
\ord_p(\om_0)\geq -1.
\ee
This means 
\be
\om_{0,p}(\pi_p\O_p)=0\qqan \om_{0,p}(\O_p)=\F_q\om_{0,p}(1).
\ee
This means that $\ord_p(\om)\geq 0$ iff $\ord_p(f)\geq 1$. This implies that
\be
-1\leq \ord_p(\om_p)\qqan |H|\cap|D|=\emptyset.
\ee
On the other hand, if $\ord_p(\om_0)\geq 0$, then $\om_0(\O_p)=0$. This would implies 
that $f\om_0(\O_p)\equiv 0$ since $f(p)$ is well-defined. This cannot happen since we 
assume that our code is not trivial. All this then implies the following
\begin{lem} The rational differential $\om_0$ should satisfy 
\be
\ord_p(\om_0)=-1.
\ee
\end{lem}
By twisting a certain $D$-unit, we always can assume that $\om_0(1)=1$. 
\begin{defn} 
A non-zero rational differential $\om_0$ is call {\it $D$-special} if 
\be
\ord_p(\om_0)=-1\qqan \om_{0,p}(1)=1.
\ee
\end{defn}

\begin{lem}\label{agal1}
There always exists non-trivial $D$-special rational differentials.
\end{lem}
\bp
It suffices to prove the existence of rational differential $\om_0$ such that 
$\ord_p(\om_0)=-1$ for all $D$. To see this, we first note that, by the Riemann-Roch 
theorem, $h^0(X,K_X(D-E))-h^0(X,K_X(-E)))=\deg(K_X(D))-(g-1)-(g-1)=n$. Therefore
by the duality theorem, within the exact sequence
\be
0\to \Omega_F(E)\buildrel{\iota_{D,E}}\over\to \Omega_F(E-D)\to 
\mathrm{Coker}(\iota_{D,E})\to 0,
\ee
the quotient space $\mathrm{Coker}(\iota_{D,E})$ is not trivial.
\ep

From now on, $\om_0$ is always taken to be  $D$-special.

Now we are ready to continue the calculation in \eqref{aeq6}.
That is, for $f\in H^0(X,\O_X(H))$ and $p\in |D|$, 
\bea
\om_p(1)\,\F_q=&\om_p(\O_p)=(f\om_0)(\O_p)\\
=&f(p) \cdot \om_0(\O_p)=f(p)\cdot \F_q\om_{0.p}(1)\\
=&f(p)\,\F_q\qquad\forall p\in |D|.
\eea
In particular, we have
\be
\om_p(1)=f(p)\qquad\forall p\in D.
\ee
This then implies the following
\begin{prop}\label{apro1}
There is a natural identification
\be
C_\Omega(D,E)=C_L(D,H).
\ee
In particular, all $\Omega$-codes are AG codes.
\end{prop}

\subsection{The duality between $C_L(D,E)$ and $\Omega_F(D,E)$}
Induced from the  natural dual pairing 
\be
\bm 
\lan\cdot,\cdot\ran_{\om_0}:&\A\times\A&\lra&\F_q\\
&(a,b)&\longmapsto&\sum_p\omega_{0,p}(a_pb_p).
\em
\ee
It is well-known that $F^\perp =F$, or equivalently,
\be\label{aeq-1}
\om_0(h)=0\qquad\forall h\in F.
\ee
Surely, this is the famous residue formula.

Now we consider the non-degenerating pairing
\be
\bm
\lan\cdot,\cdot\ran:&\F_q^n\times\F_q^n&\lra&\F_q\\
&(x,y)&\longmapsto&\sum_{i=1}^nx_iy_i.
\em
\ee
Under this pairing, we consider the image of  $C_L(D,E)\times \Omega_F(D,E)$.
\be
\bm \lan\cdot,\cdot\ran:&C_L(D,E)\times \Omega_F(D,E)&\lra&\F_q\\
&\big((f(p_1),\ldots,f(p_n)),(\om_{p_1}(1),\ldots, 
\om_{p_1}(1))\big)&\longmapsto&\sum_{i=1}^nf(p_i)\, \om_{p_i}(1).
\em
\ee
By definition, note that $\ord_{p_i}(\om/\om_0)\geq 0$, we have
\be
\sum_{i=1}^nf(p_i)\, \om_{p_i}(1)=\sum_{i=1}^n \om_{p_i}(f(p_i))=\om(f)
=\om_0(\om/\om_0 \cdot f)=0.
\ee
by \eqref{aeq-1}. This then proves the following:
\begin{prop} We have
\be
C_L(D,E)^\perp=\Omega_F(D,E).
\ee
\end{prop}

\subsection{Invariants of AG Codes $C_L(D,E)$ and $\Omega_F(D,E)$}
By the exact sequences \eqref{req-2} and \eqref{req-3}, we have
\bea
\dim_{\F_q}C_L(D,E)=&h^0(X,\O_X(E))-h^0(X,\O_X(E-D))\\
\dim_{\F_q}\Omega_F(D,E)=&h^1(X,\O_X(E-D))-h^1(X,\O_X(E)).
\eea
This implies, from the Riemann-Roch theorem, that
\bea
&\dim_{\F_q}C_L(D,E)+\dim_{\F_q}\Omega_F(D,E)\\
&=\chi(X,\O_X(E))-\chi(X,\O_X(E-D))=\deg(D)\\
&=n.
\eea
This characterizes the dimensions of the codes spaces $C_L(D,E)$ and 
$\Omega_F(D,E)$. 

To see the weights of them, we first recall that the weight $d$ of  codes $C$ is the 
biggest number such that $w(a)\geq d$ for all codewords $a\in C$. That is to say, $d$ is 
the biggest (natural) number such that for any codeword $a$, the number of its non-zero  
components is at least $d$.

Let $d_{D,E}$ be the weight of $C_L(D,E)$ which we assume to be non-trivial. 
By definition, we may find $f\in H^0(X\O_X(E))$ such that
$\wt \big(\phi_{D,E}(f)\big) = d_{D,E}$. This means that there are exactly $n-d_{D,E}$
points $p_{i_1},\ldots, p_{i_{n-d}}\in |D|$ such that $f(p_{i_j})=0$ for 
$j=1,\ldots,n-d_{D,E}$. This implies that 
$f\in H^0(X,\O_X(E-\sum_{j=1}^{n-d_{D,E}}p_{i_j})).$ In particular,
\be
0\leq \deg(E-\sum_{j=1}^{n-d_{D,E}}p_{i_j})=\deg(E)-n+d_{D,E}.
\ee
This the implies the following
\begin{prop}
\begin{enumerate}
\item [(1)] The invariants of $C_L(D,E)$ is given by
\be
\l(n, h^0(X,\O_X(E))-h^0(X,\O_X(E-D)), \geq n-\deg(E)\r)
\ee
\item [(1)] The invariants of $\Omega_F(D,E)$ is given by
\be
\l(n, h^1(X,\O_X(E-D))-h^1(X,\O_X(E)), \geq \deg(E)-2g+2\r)
\ee
\end{enumerate}
\end{prop}
\bp
It suffices to prove (2). The statements for the length and the dimension are obvious. To 
see the lower bound for the weight, we use Proposiion\,\ref{apro1}. Hence the proof for 
(1) before the proposition implies that the minimal distance of the codes 
$\Omega_F(D,E)$ is $\geq n-\deg(H)=n-\deg((\om_0)+E-n)=\deg(E)-2g-2.$
\ep

For linear codes of types $(n,k,d)$, the so-called {\it Singleton Bound}  is refer to the 
condition
\be
n+1\geq k+d.
\ee
In terms of AG codes $C_L(D,E)$ and $\Omega_F(D,E)$,
this is equivalent to
\bea\label{eqa1}
n+1\geq& h^0(X,\O_X(E))-h^0(X,\O_X(E-D))+d_{D,E}\\
\geq &\Big(h^0(X,\O_X(E))-\deg(E)\Big)-h^0(X,\O_X(E-D))+n
\eea
and
\bea\label{eqa2}
n+1\geq& h^1(X,\O_X(E-D))-h^1(X,\O_X(E))+d_{D,E}'\\
\geq &h^0(X,K_X(D-E))-h^0(X,K_X(-E))+\deg(E)-2g+2
\eea
respectively.

(1) Assume that $\deg(E-D)\leq 0$, i.e. $\deg(E)<n$, then, by the vanishing theorem
$h^0(X,\O_X(E-D)=\{0\}$ and \eqref{eqa1} becomes
\be
n+1\geq h^0(X,\O_X(E))+d_{D,E}
\geq \Big(h^0(X,\O_X(E))-\deg(E)\Big)+n\geq n-g+1
\ee
In particular, if $\deg(E)>2g-2$, then 
\be
k+d_{D,E}=n-g+1.
\ee

(2) Assume that $\deg(E)>2g-2$, then $h^1(X, \O_X(E))=\{0\}$, and  \eqref{eqa2} 
becomes
\bea
n+1\geq& h^1(X,\O_X(E-D))+d_{D,E}'\\
\geq &h^0(X,K_X(D-E))+\deg(E)-2g+2
\eea
This implies that 
\be
k_{D,E}^\perp=h^0(X,K_X(D-E))\geq (2g-2)+n-\deg(E)-(g-1)=n+g-1-\deg(E). 
\ee
In particular, if $\deg(E)<n$,
then by the discussion using $H$, we get
\be
k_{D,E}^\perp= n + g - 1 - \deg (E).
\ee
All these are surely nothing but the discussions on the invariants 
$(n,k_{D,E},d_{D,E})$ and $(n,k_{D,E}^\perp,d_{D,E}^\perp)$ for the AG codes
 $C_L(D,E)$ and $\Omega_F(D,E)$, respectively, in~\cite{S}. 

In the main text, we will introduce a high rank version for the codes above guided by the 
discussion in this appendix.
\end{appendices}

~\vskip 0.30cm

Lin WENG

Faculty of Mathematics

Kyushu University, 

Fukuoka, 819-0395, 

JAPAN

E-Mail: weng@math.kyushu-u.ac.jp
\vskip 1.0cm
\noindent
\hskip 7.80cm  (June 12, 2018 @ Fukuoka)

\end{document}